\documentclass[aps,twocolumn,showpacs,amsmath,amssymb,10pt]{revtex4-1}
\usepackage{graphicx,xcolor}
\usepackage{scrextend}
\usepackage{upgreek}
\usepackage{manfnt,pifont}
\usepackage{hyperref}
\usepackage{amsfonts}
\usepackage{tikz}

\makeatletter
\renewcommand{\p@subsection}{}
\renewcommand{\p@subsubsection}{}
\makeatother

\newcommand\noi{\noindent}

\newcommand{\be}{\begin{equation}}
\newcommand{\ee}{\end{equation}}
\newcommand{\bea}{\begin{eqnarray}}
\newcommand{\eea}{\end{eqnarray}}
\newcommand\restr[2]{{\left.\kern-\nulldelimiterspace#1\vphantom{\big|}\right|_{#2}}}

\newcommand{\sX}{\mathfrak{X}}

\newcommand{\rep}{\mathcal{R}}
\newcommand{\weight}{\omega}
\newcommand{\myzeta}{\zeta}
\newcommand{\Cas}{\mathfrak{C}}
\newcommand{\DeltaExt}{\Delta_{\mathtt{ext}}}
\newcommand{\gshort}{\mathfrak{f}}
\newcommand{\glong}{\mathfrak{F}}
\newcommand{\ffcorrshort}{f}
\newcommand{\ffcorrlong}{F}
\newcommand{\ext}{\mathtt{ext}}
\newcommand{\metric}{\mathsf{g}}
\newcommand{\JJ}{j^2_{\Delta}}
\newcommand{\GammaC}{\mathsf{X}}

\begin{document}
\title{
Bootstrapping the half-BPS line defect CFT in $\mathcal{N}=4$ SYM at strong coupling
}
\author{Pietro Ferrero$^{\phi}$ and Carlo Meneghelli$^{\psi,f}$ } 
\affiliation{ \mbox{$\phi$ Mathematical Institute, University of Oxford, Woodstock Road, Oxford, OX2 6GG, U.K.}
\\
 \mbox{$\psi$\,\,Dipartimento SMFI, Universit\`{a} di Parma, Viale G.P.
Usberti 7/A, 43100, Parma, Italy}\\
 \mbox{$f$\, INFN Gruppo Collegato di Parma}
 \\
{\tt  \mbox{pietro.ferrero$\bullet$maths.ox.ac.uk, carlo.meneghelli$\bullet$gmail.com}}}
\begin{abstract}
\noi 
We consider the one-dimensional (1d) CFT defined by the half-BPS Wilson line in planar $\mathcal{N}=4$ super Yang-Mills.
Using analytic bootstrap methods we derive the four-point function of the super-displacement operator at fourth order in a strong coupling expansion. Via AdS/CFT, this corresponds to the first three-loop correlator in anti-de-Sitter ever computed. To do so we address the operator mixing problem by considering a family of auxiliary correlators. We further extract the anomalous dimension of the lightest non-protected operator and find agreement with the integrability-based numerical result of Grabner, Gromov and Julius. 
\end{abstract}
\maketitle

\section{Introduction}\label{sec:intro}
It is hard to overstate the fundamental role played by symmetries and consistency conditions in quantum field theory (QFT).
This is especially true for conformal field theories (CFTs) where the latter can be explicitly formulated and used to give concrete predictions for observable quantities.
This strategy, called the conformal bootstrap, has produced spectacular results over the past decade or so (see e.g.~\cite{Poland:2018epd}).

Here, we focus on a one-parameter family of one-dimensional (1d) CFTs \cite{Qiao:2017xif} with extended supersymmetry, namely $\mathfrak{osp}(4^*|4)$,  that admits two holographically dual realizations: 
as a line defect in planar $d=4$,  $\mathcal{N}=4$ super Yang-Mills, namely as a Wilson line in the fundamental representation, or as a two-dimensional QFT in $\text{AdS}_2$ \cite{Giombi:2017cqn}.  This family of CFTs is parametrized by the ’t~Hooft coupling $\lambda$ and each description is perturbative in opposite regimes. 

The aim of this paper is to show  that this 1d CFT can be efficiently and systematically solved perturbatively at strong coupling (large $\lambda$) using analytic bootstrap methods introduced in 1d in \cite{FernandoNotes,Liendo:2018ukf,Ferrero:2019luz}.  The power and success of our procedure is established by the determination of the scaling dimension of the lightest non-protected scalar operator to be \footnote{The definition of $\lambda$ is fixed by the OPE coefficient \eqref{ShortOPE}.}
\begin{equation}
\label{Delta}
\Delta_{\phi^2}\!=\!2-\tfrac{5}{\sqrt{\lambda}}+\tfrac{295}{24}\tfrac{1}{\lambda}-\tfrac{305}{16}\tfrac{1}{\lambda^{3/2}}+ \left(\tfrac{351845}{13824}-\tfrac{75}{2}\zeta(3)\right)\tfrac{1}{\lambda^{2}}+\dots .
\end{equation}
Excitingly, this formula agrees with the numerical result obtained in \cite{Grabner:2020nis} by the completely independent, integrability-based,  quantum spectral curve method. 
We extract \eqref{Delta} from the four-point function of the so-called super-displacement operator, which we bootstrap up to the same order. Its explicit expression is given in an attached notebook.  With the current technology it would have been impossible to determine this correlator directly from Witten diagrams as it corresponds to a three-loop computation in anti-de-Sitter (AdS).

The basic idea, applied in e.g.~\cite{Aharony:2016dwx,Aprile:2017bgs,Alday:2017zzv,Caron-Huot:2020bkp}, is to construct an ansatz for the correlator and to impose consistency with the operator product expansion and Bose symmetry.  
The main obstacle in implementing this procedure is the problem of operator mixing.
This makes it necessary to consider not just one,  but a whole family of correlators that is large enough depending on the specific CFT and on the perturbative order one is interested in.
An interesting feature of this 1d CFT is that the degeneracies in the spectrum of conformal dimensions at $\lambda=\infty$ are unaffected at the first perturbative order. Thus,  mixing plays a noticeable role in the determination of the four-point function of the super-displacement operator starting at fourth order, when the square of the second order anomalous dimension matrix first appears.
Additionally, the knowledge of all four-point functions of half-BPS operators at second order,  which will be presented in \cite{Longversion}, is not enough to take into account this effect and correlators involving non-protected external operators need to be included. This goes fundamentally beyond what has been done in previous works, for example \cite{Aprile:2017bgs,Alday:2017xua,Aprile:2017bgs,Alday:2020tgi}.

\section{Superconformal symmetry} \label{sec:symmetry}

The symmetry of the 1d super-CFT (SCFT) we are studying is $\mathfrak{osp}(4^*|4)$.  Its bosonic subalgebra is $\mathfrak{so}(4^*)\oplus \mathfrak{sp}(4)\simeq \mathfrak{sl}(2)\oplus \mathfrak{su}(2)\oplus \mathfrak{sp}(4)$, where the first term corresponds to the 1d conformal group while the remaining two can be thought of as R symmetries.
The relevant representations of $\mathfrak{osp}(4^*|4)$ are uniquely specified by the scaling dimensions and R-symmetry representation $\weight=\{\Delta,s,[a,b]\}$ \footnote{
Here $s\in \mathbb{N}$ corresponds to the $s+1$ dimensional representation of $\mathfrak{su}(2)$  while $[a,b]$ are $ \mathfrak{sp}(4)$ Dynkin labels, so that $[1,0]=\mathbf{4}$ and $[0,1]=\mathbf{5}$.}
 of the superconformal primary.  
 Two types of supermultiplets $\mathcal{R}$ will be relevant in this work:
 (i)  long multiplets  $\mathcal{L}_{s,[a,b]}^{\Delta}$ where $\Delta$ is subject to the unitarity bound,
(ii) short multiplets  $\mathcal{D}_k$, with $\weight=\{k,0,[0,k]\}$.
A distinguished role is played by the super-displacement operator $\mathcal{D}_1$,  which is ultra short and whose decomposition in irreducible representations of the bosonic symmetry is
\begin{equation}
\label{D1multiplet}
\mathcal{D}_1:\,\,\,\phi^{\Delta=1}_{(\mathbf{1},\mathbf{5})}\,
\rightarrow\,
\psi^{\Delta=3/2}_{(\mathbf{2},\mathbf{4})}\,
\rightarrow\,
f^{\Delta=2}_{(\mathbf{3},\mathbf{1})}\,,
\end{equation}
where the arrow refers to the action of supersymmetry generators,  while $(\mathbf{m},\mathbf{n})$ denotes the dimensions of the $\mathfrak{su}(2)\oplus \mathfrak{sp}(4)$ representation.
In the following  we will consider four-point functions of two types 
\begin{equation}
\label{3typsofcorrelators}
\langle \mathcal{D}_{1} \mathcal{D}_{1} \mathcal{D}_{k} \mathcal{D}_{k}\rangle\,,
\quad \,\,\,\,
\langle \mathcal{D}_{1} \mathcal{D}_{1} \mathcal{D}_{2} \,\mathcal{L}_{0,[0,0]}^{\DeltaExt}\rangle\,.
\end{equation}
The implications of superconformal symmetry on correlation functions involving only short operators have been analyzed in
 \cite{Liendo:2016ymz,Liendo:2018ukf} using superspace.  They not only imply that the four-point functions of all the members of the short supermultiplet are determined by one of the superprimaries, but that the latter are subject to constraints. In the simplest example of
  $\langle \mathcal{D}_{1} \mathcal{D}_{1} \mathcal{D}_{1} \mathcal{D}_{1}\rangle$ 
these can be solved in terms of a constant  and a single function of the bosonic cross ratio $\chi=\tfrac{t_{12}t_{34}}{t_{13} t_{24}}$, where $t$ is a coordinate on the line and $t_{ij}=t_i-t_j$. The explicit parametrization is
\begin{equation}
\label{D1fourpoints}
\frac{\langle \mathcal{D}_{1} \mathcal{D}_{1} \mathcal{D}_{1} \mathcal{D}_{1}\rangle}{\langle \mathcal{D}_{1} \mathcal{D}_{1}\rangle \langle \mathcal{D}_{1} \mathcal{D}_{1}\rangle}\,=
\mathsf{f}\,\sX+\mathbb{D}\ffcorrshort(\chi)\,,
\end{equation}
where the super-conformal invariant $\sX$ and the differential operator $\mathbb{D}$ are given in appendix \ref{app:blockology}.
The number $\mathsf{f}$ in \eqref{D1fourpoints} is a datum of the topological algebra associated with any 1d CFT with $\mathfrak{osp}(4^*|4)$ symmetry by the cohomological construction of  \cite{Chester:2014mea,Beem:2016cbd}; see \cite{Liendo:2016ymz}. If the 1d CFT in question is a Wilson line in $\mathcal{N}=4$ SYM, $\mathsf{f}$ can be computed by supersymmetric localization \cite{Pestun:2007rz,Giombi:2009ds,Giombi:2018qox}.
See \cite{Liendo:2018ukf}  for more details.

To  address the mixing problem we also consider correlators of the second type in \eqref{3typsofcorrelators}. Superconformal symmetry implies that each of them is determined by a single function $\ffcorrlong(\chi)$; see appendix \ref{app:blockology} and \cite{Longversion}.

Correlation functions of local operators admit a decomposition in superconformal blocks,  defined by the $\mathfrak{osp}(4^*|4)$ Casimir equation supplemented with the appropriate boundary conditions.  We parametrize the Casimir eigenvalues as
\begin{equation}
\label{Casimir}
\Cas_2(\rep)=
\Delta(\Delta +3)+\tfrac{ s(s+2)}{4} -\tfrac{a^2}{2}-a (b+2)-b (b+3)\,,
\end{equation}
with $\weight_{\rep}=\{\Delta,s,[a,b]\}$.
Explicit expressions of superconformal blocks are given in appendix \ref{app:blockology},  and their derivation will be presented in \cite{Longversion}.
 
The conformal blocks decomposition of any four-point function follows from the OPE rules.  In the case of $\mathcal{D}_1$ they take the form
 \begin{equation}
 \label{D1D1OPE}
 \mathcal{D}_1\times  \mathcal{D}_1 = \mathcal{I}+ \mathcal{D}_2+\mathcal{L}^{\Delta}_{0,[0,0]}\,.
  \end{equation}
The OPE $ \mathcal{D}_k\times  \mathcal{D}_k$ has the same form plus extra representations that are projected away in the correlator of interest.
 It follows that the decomposition of \eqref{D1fourpoints} in superconformal blocks is
  \begin{equation}
\label{ConfBlockdecompositionGeneralD1D1D1D1}
\ffcorrshort(\chi)=
\gshort_{\mathcal{I}}(\chi)+ \mu^2_{\mathcal{D}_2}\,\gshort_{\mathcal{D}_2}(\chi)
+\sum_{\mathcal{O}}\, \mu^2_{\mathcal{O}}\,\gshort_{\Delta_{\mathcal{O}}}(\chi)\,,
\end{equation}
and $\mathsf{f}=1+ \mu^2_{\mathcal{D}_2}$, where $\mathcal{O}$ are superconformal primaries of type $\mathcal{L}^{\Delta}_{0,[0,0]}$.
We shall also use the OPE\footnote{
There is a unique super-conformal invariant structure  of type $\langle  \mathcal{D}_2\, \mathcal{L}^{\DeltaExt}_{0,[0,0]}\mathcal{L}^{\Delta}_{0,[0,0]} \rangle$. The relevant OPE coefficient can be extracted from the three-point correlator of the $\mathcal{D}_2$ and  $\mathcal{L}^{\DeltaExt}_{0,[0,0]}$ superprimaries with the $Q^{4}|_{0,[0,2]}$ descendant of the exchanged operator of type $\mathcal{L}^{\Delta}_{0,[0,0]}$.
}
 \begin{equation}
 \label{D2longOPE}
  \mathcal{D}_2\times \mathcal{L}^{\DeltaExt}_{0,[0,0]} = \mathcal{D}_2+\mathcal{L}^{\Delta}_{0,[0,0]}+\dots\,\,\,,
 \end{equation}
 where $\dots$ indicates representations that do not contribute to \eqref{3typsofcorrelators}. The selection rules for the other channel can be found in appendix \ref{app:blockology}.

\section{Free theory}\label{sec:free}
The free 1d CFT from which we start the perturbation is easy to describe.  Its local operators are built by taking normal ordered products of the fundamental fields $\Phi=(\phi^{I}(t),\psi^A_{\alpha}(t),f_{\alpha \beta}(t))$, 
with $I=1,\dots,5$, $A=1,\dots,4$, $\alpha,\beta=1,2$ (see \eqref{D1multiplet}) and their derivatives. Correlation functions are defined and computed by Wick contractions using the two-point function of $\Phi$. By the state-operator correspondence we can think in terms of the space of states
\begin{equation} 
\label{Hdecomposition}
\mathcal{H}= \bigoplus_L \mathcal{H}_L\,
\qquad
 \mathcal{H}_L=\left(\mathbb{V}_{\Phi} \otimes \dots \otimes \mathbb{V}_{\Phi}\right)^{S_L},
\end{equation}
where $\mathbb{V}_{\Phi}\simeq \mathcal{D}_1$, the symbol $S_L$ indicates that the tensor products are totally graded-symmetrized and the integer $L$ corresponds to the length of composite operators.  Each factor $ \mathcal{H}_L$ decomposes into irreducible representations of $\mathfrak{osp}(4^*|4)$ as
\begin{equation} 
\label{HLdecomposition}
 \mathcal{H}_L= \bigoplus_\mathcal{R}\, \mathsf{d}_{L}(\rep)\otimes \rep\,,
 \end{equation}
 where  $\mathsf{d}_{L}(\rep)$ are multiplicity spaces. 
 Their dimensions can be obtained  by expanding the partition function that counts the words made of  $\Phi$ and its derivatives in characters of $\mathfrak{osp}(4^*|4)$ (see \cite{Liendo:2018ukf}).
For length two, the decomposition \eqref{HLdecomposition} is multiplicity free and contains only the multiplets $\mathcal{D}_2$ and  $\mathcal{L}_{0,[0,0]}^{\Delta}$ with $\Delta=2,4,\dots$.
After turning on the perturbation, the corresponding operators will mix with operators in the same representation in $\mathcal{H}_{L>2}$. 
At the perturbative order considered in this work a fundamental role will be played by length four operators in such representations; their number is given by
\begin{equation}\label{quadraticdegeneracy}
\text{dim}\left(\mathsf{d}_4(
\Delta
)
\right)=\mathtt{Floor}\left[  \left(\tfrac{\Delta}{4}\right)^2  \right]\,,
\,\,\, \Delta=4,6,\dots\,\,,
\end{equation}
where we introduced the notation  $\mathsf{d}_{L}(\Delta):= \mathsf{d}_{L}(\mathcal{L}_{0,[0,0]}^{\Delta})$.  While this counting gives valuable information,  for our purposes we will need to construct the length four operators explicitly.

The finite dimensional multiplicity spaces $\mathsf{d}_{L}(\rep)$ are equipped with an inner product $\metric$, which is determined by the two-point functions in the free theory and does not mix operators of different lengths. Additionally, three-point functions provide trilinear maps
 \begin{equation}
 \label{CtrilinearMAPS}
 \mathsf{C}^{(0)}:\,
 \mathsf{d}_{L_1}(\rep_1)
 \times \mathsf{d}_{L_2}(\rep_2)
  \times \mathsf{d}_{L_3}(\rep_3)\rightarrow \mathbb{C}^{\#}\,,
 \end{equation}
 where $\#$ denotes the number of invariant structures of type $\langle \rep_1\rep_2\rep_3\rangle$.  Only situations with $\#=1$ will be relevant in this work.

\section{The bootstrap problem} \label{sec:bootstrap}

\subsection*{The mixing problem}
 Consider the conformal block decomposition \eqref{ConfBlockdecompositionGeneralD1D1D1D1} and expand the CFT data in a small parameter $1/\sqrt{\lambda}$, for example
\begin{equation}\label{anomalousdimensions}
\Delta_{\mathcal{O}}^{}=
\Delta_{\mathcal{O}}^{(0)}+
\tfrac{1}{\sqrt{\lambda}}
\gamma^{(1)}_{\mathcal{O}}+
\tfrac{1}{\lambda}
\gamma^{(2)}_{\mathcal{O}}+\dots\,\,,
\end{equation}
and similarly for the OPE coefficients. This produces logarithms in the small $\chi$ expansion of the correlation function. More precisely, the correlator at order $\ell$ has the structure
  \begin{equation}
  \label{fwithLOGS}
\ffcorrshort^{(\ell)}(\chi)=
\sum_{k=0}^{\ell}
\ffcorrshort^{(\ell)}_{\log^{k}}(\chi)\left(\log \chi\right)^{k}\,,
\end{equation}
where $\ffcorrshort^{(\ell)}_{\log^{k}}(\chi)$ are analytic at $\chi=0$.
Their explicit expression in terms of CFT data is given in appendix \ref{app:blockology}.
The functions that multiply higher powers of the logarithms (those with $k>1$) are expressed in terms of CFT data at lower order. What makes the bootstrap problem more complicated, but also more interesting, is that, in general,  due to degeneracies in the spectrum of the free theory, these CFT data cannot be obtained from the correlator \eqref{D1fourpoints} alone.

From the knowledge of $\ffcorrshort^{(0)}(\chi)$  and $\ffcorrshort^{(1)}(\chi)$  one can extract, via the decomposition \eqref{ConfBlockdecompositionGeneralD1D1D1D1},  the combinations
\begin{align}
\label{aaverage}
\langle a^{(0)}_{\Delta}\rangle\,&:=\,\sum_{\mathcal{O} |\Delta^{(0)}_{\mathcal{O}}=\Delta}\,(\mu^{(0)}_{\mathcal{O}})^2\,,\\
\label{gammaaaverage}
\langle a^{(0)}_{\Delta}\gamma^{(1)}_{\Delta}\rangle\,:&=\,\sum_{\mathcal{O} |\Delta_{\mathcal{O}}^{(0)}=\Delta}\,(\mu^{(0)}_{\mathcal{O}})^2\,\gamma^{(1)}_{\mathcal{O}}\,,
\end{align}
where $\Delta=2,4,6,\dots$.
To reconstruct the highest logarithm at the next order, namely $\ffcorrshort^{(2)}_{\log^{2}}(\chi)$, one needs to know the quantity
 $\langle a^{(0)}_{\Delta}(\gamma^{(1)}_{\Delta})^2\rangle$, but these  ``averaged moments'' cannot be extracted from \eqref{aaverage} and \eqref{gammaaaverage} when operators are degenerate.
The 1d SCFT  analyzed here has the following interesting property:  at first order, the anomalous dimension of any operator is proportional to the eigenvalue of the  quadratic Casimir of $\mathfrak{osp}(4^*|4)$ \eqref{Casimir}
\begin{equation} \label{treegammas}
\gamma^{(1)}_{\mathcal{O}}\,=\,-\tfrac{1}{2}\,\Cas_2(\rep_{\mathcal{O}})\,.
\end{equation}
This implies that the degeneracy is not lifted at first order,  and hence any factor of
$\gamma^{(1)}_{\mathcal{O}}$ in averages of the type \eqref{gammaaaverage}  can be replaced by \eqref{treegammas} and pulled out of the sum. 
A simple proof of \eqref{treegammas} based on the properties of the dilatation operator, which can be derived from Witten diagram considerations or directly from the bootstrap, is presented in appendix \ref{app:treegammas}. More details will be presented in \cite{Longversion}.

Let us move to higher orders in the perturbative expansion.  By looking at the expression of $\ffcorrshort^{(\ell)}_{\log^{k}}(\chi)$ entering \eqref{fwithLOGS} in terms of CFT data, it is not hard to realize that the first time an unknown combination of CFT data appears for the higher logarithms $k>1$ is at fourth order.  Specifically, $\ffcorrshort^{(4)}_{\log^{2}}(\chi)$ contains terms of the form
\begin{equation}
\label{gamma2firsttime}
\langle a^{(0)}_{\Delta}(\gamma^{(2)}_{\Delta})^2\rangle\,:=\,\sum_{\mathcal{O} |\Delta^{(0)}_{\mathcal{O}}=\Delta}\,(\mu^{(0)}_{\mathcal{O}})^2\,(\gamma^{(2)}_{\mathcal{O}})^2\,.
\end{equation}
The main obstacle to bootstrap the correlator \eqref{D1fourpoints} at this order is to determine \eqref{gamma2firsttime}.
Luckily, to do this we do not need to find the eigenvalues $\gamma^{(2)}_{\mathcal{O}}$ and eigenvectors $\mathcal{O}$ of the second order dilatation operator.  
In fact, we can work in an arbitrary (non-orthogonal) basis  for the exchanged operators.  It is convenient to use a basis in which  the length $L$ is a good quantum number, which implies two simplifications.  The first is obvious:  the three-point function $\mathsf{C}^{(0)}_{\mathcal{D}_1 \mathcal{D}_1 -}$ (see \eqref{CtrilinearMAPS}) is non vanishing only if the third operator has length two -- these are non degenerate and we denote the corresponding OPE coefficient by
$\mathsf{C}^{(0)}_{11\Delta}$. 
The second concerns the anomalous dimension matrix  $\Gamma^{(2)}_{\Delta}$: its components are non vanishing only among operators of the same length or whose lengths differ by two units \footnote{
This can be argued from the structure of the Witten diagrams or directly from the bootstrap, see \cite{Longversion}.}.
We denote the corresponding building blocks by $\Gamma^{(2)}_{\Delta,\,L\rightarrow L}$ and $\Gamma^{(2)}_{\Delta,\,L\rightarrow L+2}$,  with 
\begin{equation}
\label{Gamma2belongsto}
\Gamma^{(2)}_{\Delta,\,L_1\rightarrow L_2}\in \mathsf{d}_{L_1}(\Delta)\times  \mathsf{d}_{L_2}(\Delta)\,,
\end{equation}
where the degeneracy spaces $\mathsf{d}_{L}(\Delta)$ were defined in \eqref{HLdecomposition}.  In appendix \ref{app:1111} we give explicit examples for $\Delta=4,6$. We conclude that in this basis, after normalizing the length-two operators,  the expression \eqref{gamma2firsttime} takes the form
 \begin{align}
 \label{gamma2secondttime}
\langle a^{(0)}_{\Delta}(\gamma^{(2)}_{\Delta})^2\rangle=\langle a^{(0)}_{\Delta}\rangle\,\left[(\Gamma^{(2)}_{\Delta,2\to 2})^2+\delta \Gamma_{\mathtt{sq}}^{(2)}(\Delta)\right]\,,
\end{align}
where
\begin{equation}
\label{deltagamma2secondttime}
\delta \Gamma_{\mathtt{sq}}^{(2)}(\Delta)\,:=\,
\Gamma^{(2)}_{\Delta,\,2 \rightarrow 4}\cdot\,
 \metric^{}_4
\,\cdot\,
\Gamma^{(2)}_{\Delta,\,2 \rightarrow 4}\,,
\end{equation}
and $ \metric^{}_4$ is  the metric in the space 
\eqref{quadraticdegeneracy}.
While the number $\Gamma^{(2)}_{\Delta,\,2 \rightarrow 2}$ is obtained from $\ffcorrshort^{(2)}(\chi)$, to extract the vector $\Gamma^{(2)}_{\Delta,\,2 \rightarrow 4}$ one has to consider a family of correlators at second order.
A natural choice is given by
\begin{align}
\label{AuxilliaryCORRE}
\langle  \mathcal{D}_1\, \mathcal{D}_1\, \mathcal{D}_2\,\mathcal{O}_{\mathtt{ext}}\rangle\,,
\end{align}
with $\mathcal{O}_{\mathtt{ext}}\!$  either of type $\mathcal{D}_2$ or $\mathcal{L}^{\DeltaExt}_{0,[0,0]}$. 
From the superconformal block decomposition of the correlator \eqref{AuxilliaryCORRE} in the channel
\eqref{D1D1OPE}
and
\eqref{D2longOPE} 
we can extract 
\begin{align}
 \label{avggamma2long}
\langle a^{(0)}_{\Delta} \gamma^{(2)}_{\Delta}\rangle^{}_{112\mathcal{O}_{\mathtt{ext}}}\!\!=\!
\langle a^{(0)}_{\Delta}\rangle^{}_{112\mathcal{O}_{\mathtt{ext}}}\,
\Gamma^{(2)}_{\Delta,\,2 \rightarrow 2}
+
\mathsf{C}^{(0)}_{11\Delta}\,
\,\GammaC_{\Delta,\mathcal{O}_{\mathtt{ext}}}
\,,
\end{align}
where \footnote{
According to \eqref{CtrilinearMAPS}, upon specifying the second and third operator in $\mathsf{C}^{(0)}$ we obtain a map
 $\mathsf{C}^{(0)}_{ \mathcal{D}_2\mathcal{O}_{\mathtt{ext}}}: \mathsf{d}_4(\Delta)\rightarrow \mathbb{C}$. This map is composed with $\Gamma^{(2)}_{\Delta,\,2 \rightarrow 4}$, see \eqref{Gamma2belongsto}, to produce a number.
}
\begin{equation}
\label{C0dotGamma2}
\GammaC_{\Delta,\mathcal{O}_{\mathtt{ext}}}:=\Gamma^{(2)}_{\Delta,\,2 \rightarrow 4}\, \cdot  \mathsf{C}^{(0)}_{ \mathcal{D}_2\mathcal{O}_{\mathtt{ext}}}\,.
\end{equation}
The subscript in $\langle \cdots\rangle^{}_{112\mathcal{O}_{\mathtt{ext}}}$ entering \eqref{avggamma2long} indicates that the average in question is determined by the correlator \eqref{AuxilliaryCORRE}, in contrast with all the averages  $\langle \cdots\rangle$ encountered so far which correspond to the four-point function \eqref{D1fourpoints}. 

To extract $ \Gamma^{(2)}_{\Delta,\,2 \rightarrow 4}$ from \eqref{C0dotGamma2} we need to consider enough external operators  such that the vectors $\mathsf{C}^{(0)}_{\mathcal{D}_2\mathcal{O}_{\mathtt{ext}}}$
span, upon varying $\mathcal{O}_{\mathtt{ext}}$,  the whole degeneracy space $\mathsf{d}_4(\Delta)$. 
As their dimension grows quadratically with $\Delta$ \footnote{This is an important difference compared to \cite{Alday:2017xua,Aprile:2017bgs,Alday:2020tgi}. } (see \eqref{quadraticdegeneracy}),  the number of auxiliary correlators \eqref{AuxilliaryCORRE} that we consider should grow accordingly. 
This is achieved by taking as external operator $\mathcal{D}_2$ together with all the $L=2$ and $L=4$ operators of type $\mathcal{L}^{\DeltaExt}_{0,[0,0]}$ where $\DeltaExt=2,4,\dots, $ up to some maximum depending on the value of $\Delta$ in $\Gamma^{(2)}_{\Delta,\,2 \rightarrow 4}$ \footnote{With the maximum values $\DeltaExt=14$ and  $\DeltaExt=22$ for the length-four and length-two operators  one can span all directions of $\mathsf{d}_4(\Delta)$ up to $\Delta=26$.}. 
The three-point functions  in \eqref{C0dotGamma2} are a crucial input for this procedure. 
We compute them in the free theory after constructing the operators explicitly using a new method described in \cite{Longversion}.
Notice that $\mathcal{D}_k\times \mathcal{D}_k$ probes the same direction in $\mathsf{d}_4(\Delta)$ for any $k>1$,  so half-BPS external operators are insufficient to take mixing  into account .

\subsection*{The ansatz}

To bootstrap perturbative correlators we follow and develop the strategy introduced in \cite{Liendo:2018ukf} and extended to  higher orders  in \cite{Ferrero:2019luz} (see also \cite{Gimenez-Grau:2019hez,Bianchi:2020hsz,Drukker:2020swu}),  which uses a basis of harmonic polylogarithms.  In \cite{FernandoNotes,Ferrero:2019luz} it was argued that the correct basis for 1d CFTs contains the ``words'' that can be built using the symbol map \cite{Duhr:2011zq} from the two ``letters'' $\chi$ and $1-\chi$. 
The use of such a basis requires an external input,  namely the maximal transcendental weight $\mathtt{t}$ of the harmonic polylogarithms.  Because of the structure of the perturbative OPE and the polynomiality in $\Delta$ of the first-order anomalous dimensions,  the correct choice of basis at the $\ell$-th perturbative order has $\mathtt{t}=\ell$.  An explicit basis is given in appendix \ref{app:basis} up to $\mathtt{t}=4$; its dimension is $\sum_{\mathtt{t}=0}^{\ell}2^{\mathtt{t}}=2^{\ell+1}-1$.  
For a generic $\ell$-th order correlator $G^{(\ell)}(\chi)$,  we make the ansatz
\begin{align} \label{transcendental_ansatz}
G^{(\ell)}(\chi)=\sum_{i=1}^{2^{\ell+1}-1}r_i(\chi)\,\mathcal{T}_i(\chi)\,,
\end{align}
where $r_i(\chi)$ are polynomials in $\chi$ divided by powers of $\chi$ and $(1-\chi)$ \cite{Ferrero:2019luz}, while $\mathcal{T}_i(\chi)$ form our basis of harmonic polylogarithms for transcendentality up to $\mathtt{t}=\ell$. The bootstrap problem is then reduced to that of fixing the rational functions $r_i(\chi)$ appearing in \eqref{transcendental_ansatz}.  In the following we describe the strategy for the two types of correlators introduced in \eqref{3typsofcorrelators}.

We can fix the correlator   $\langle \mathcal{D}_1\,\mathcal{D}_1\,\mathcal{D}_1\,\mathcal{D}_1\rangle$  completely up to fourth order using the ansatz described above,   with the following constraints:
\begin{enumerate}
\item Crossing symmetry,  which for the reduced correlator $\ffcorrshort(\chi)$ appearing in \eqref{D1fourpoints} reads
\begin{align}
(1-\chi)^2\,\ffcorrshort(\chi)+\chi^2\,\ffcorrshort(1-\chi)\,=\,0\,.
\end{align}
\item As discussed,  at every order the highest powers of $\log\chi$ (those with $k>1$ in eq.  \eqref{fwithLOGS}) in the $\chi\to 0$ limit can be obtained from previous order data.  
\item The invariance of the free theory under $\chi\to \frac{\chi}{\chi-1}$ is ``weakly'' broken by perturbative corrections,  but it still constrains correlators at each order,  see \cite{Liendo:2018ukf,  Ferrero:2019luz}. 
\item As discussed in section \ref{sec:symmetry},  the quantity $\mu^2_{\mathcal{D}_2}$ is known from localization.  The first orders read
\begin{align}
\label{ShortOPE}
\mu^2_{\mathcal{D}_2}=2-\tfrac{3}{\lambda^{1/2}}+\tfrac{45}{8\,\lambda^{3/2}}+\tfrac{45}{4\,\lambda^2}+\dots\,\,\,.
\end{align}
This provides the definition of the coupling $\lambda$. 
\end{enumerate}
This fixes $f(\chi)$ at each order up to polynomial ambiguities in the anomalous dimensions, of degree $6,10,14,\dots\,$ in $\Delta$, which we fix by requiring that,  at each order, the average anomalous dimension has the mildest possible growth at large $\Delta$:
\begin{align}
\label{LargeDeltaGROWTH}
\langle \gamma^{(\ell)}_{\Delta}\rangle:=\frac{\langle a^{(0)}_{\Delta}\gamma^{(\ell)}_{\Delta}\rangle}{\langle a^{(0)}_{\Delta}\rangle}
 \sim \Delta^{\ell+1} \qquad 1\le \ell\le 4\,.
\end{align}

We now move to correlators \eqref{AuxilliaryCORRE}, with $\mathcal{O}_{\mathtt{ext}}=\mathcal{L}^{\DeltaExt}_{0,[0,0]}$.  Given the knowledge of \eqref{treegammas},  the free theory and first-order results do not present any conceptual novelty.  At second order,  it is important to realize that not all the correlator is necessary to extract the averages \eqref{avggamma2long}.  Rather,  it is enough to bootstrap the part of the correlator that produces a $\log\chi$ in the OPE around $\chi=0$,  where those averages first appear: we shall refer to this as $\ffcorrlong^{(2)}_{\log}(\chi)$.  Once the terms of transcendentality 2 in the ansatz are fixed by lower-order data and property (c),  $\ffcorrlong^{(2)}_{\log}(\chi)$ is known up to a finite number of undetermined coefficients,  which we fix with the following recursive procedure.

Imagine knowing all $\ffcorrlong^{(2)}_{\log}(\chi)$ up to $\DeltaExt=\widehat{\Delta}-2$: these allow one to extract $\Gamma^{(2)}_{\Delta,2\to 4}$ at least for $\Delta=2,\dots,\widehat{\Delta}+2$.  We use the latter to compute the combination \eqref{avggamma2long} for the same values of $\Delta$, but now averaged over each correlator with $\DeltaExt=\widehat{\Delta}$, which in turn fixes $\ffcorrlong^{(2)}_{\log}(\chi)$ completely for $\DeltaExt=\widehat{\Delta}$.  Once that is known,  we can compute new entries of $\Gamma^{(2)}_{\Delta,2\to 4}$,  that we use for the following recursive step.  The starting point of the recursion is $\DeltaExt=2$, which can be fixed by using the averages computed from $\langle \mathcal{D}_1\,\mathcal{D}_1\,\mathcal{D}_k\,\mathcal{D}_k\rangle$.

\subsection*{Results}
Our first important intermediate result is the computation of the average \eqref{gamma2secondttime},  which is necessary to bootstrap $f^{(4)}(\chi)$.  While $\Gamma^{(2)}_{\Delta,2\to 2}$ can be found in eq.  (6.24) of \cite{Liendo:2018ukf}, the newly computed contribution due to mixing is
\begin{align}
\begin{split}
\delta\Gamma_{\mathtt{sq}}^{(2)}(\Delta)=&\JJ \left(\tfrac{(\JJ-2)}{2}\,S_{-2}(\Delta)+\tfrac{3\,\JJ-4}{8}\,H_{\Delta}^2\right)\\
&-\tfrac{p_1(\Delta)}{(\Delta+1)(\Delta+2)}\,H_{\Delta}+\tfrac{p_2(\Delta)}{(\Delta+2)}\,,
\end{split}
\end{align}
where $\JJ=\Delta(\Delta+3)$,  $S_{-2}(\Delta)=\sum_{n=1}^{\Delta}\frac{(-1)^n}{n^2}$, $H_{n}$ is the $n$-th harmonic number and $p_{1,2}(\Delta)$ are polynomials given in appendix \ref{app:1111}.

Our main result is the determination of $f(\chi)$ up to fourth order at large $\lambda$. The explicit expressions are contained in a notebook included with the submission. From this correlator we extract the conformal dimension, see  \eqref{Delta}, and squared OPE coefficient of the lightest non-protected operator $\phi^2$
\begin{align}
\begin{split}
\mu^2_{\phi^2}=&\tfrac{2}{5}-\tfrac{43}{30\,\sqrt{\lambda}}+\tfrac{5}{6\,\lambda}+\left(\tfrac{11195}{1728}+4\zeta(3)\right)\tfrac{1}{\lambda^{3/2}}\\
&-\left(\tfrac{1705}{96}+\tfrac{1613}{24} \zeta (3)\right)\tfrac{1}{\lambda^2}+\dots\,\,\,.
\end{split}
\end{align}
More averaged CFT data extracted from this correlator are given in appendix \ref{app:1111}.

\section{Discussion}\label{sec:discussion}
In this paper we have shown how to bootstrap correlation functions in perturbation theory for a special 1d SCFT from  the knowledge of the unperturbed theory, symmetries, consistency conditions and some extra physical input.
To implement this program we put forward a new strategy to take into account operator degeneracies which we believe can be applied more broadly, e.g., for holographic CFTs \cite{Alday:2017xua,Aprile:2017bgs,Alday:2020tgi} at higher orders and for the $\epsilon$-expansion in \cite{Alday:2017zzv,Carmi:2020ekr}.
There are several interesting open questions for the future.

The first direction is to consider higher orders \footnote{While the perturbative expansion in $\lambda^{-1/2}$ is asymptotic, the expansion of CFT data, like  $\mu^2_{\mathcal{D}_2}$, as a function of $\Delta_{\phi^2}-2$ appears to be convergent.
} in the perturbative expansion. This is technically more challenging since additional operators will participate in the mixing, but also involves new conceptual problems related to additional ambiguities and the uniqueness of the theory; see discussion above \eqref{LargeDeltaGROWTH}. To address this question it will be useful to study the 1d SCFT defined by Wilson lines with different gauge groups, still at large rank, and in different representations; see e.g.~\cite{Gomis:2006sb,Giombi:2020kvo}.

The theory we are considering is supposed to be integrable \cite{Drukker:2006xg,Drukker:2012de,Correa:2012hh}. In this work we used the integrability-based results of  \cite{Grabner:2020nis} only as a check of our procedure. It would be interesting to numerically determine the conformal dimension of other operators in addition to \eqref{Delta} as a function of the coupling using the method of  \cite{Grabner:2020nis} and compare to our findings. 
How to directly incorporate integrability into the bootstrap remains a crucial open question, see \cite{Kiryu:2018phb} for explorations in this direction.

Finally, the 1d SCFT considered in this work is also an excellent playground to test and use the so-called inversion formula of \cite{Mazac:2018qmi}.

\section*{acknowledgments}
We thank Fernando Alday and Pedro Liendo  for collaboration at early stages of this work and for many fruitful discussions. We thank Fernando Alday for sharing with us unpublished notes and are grateful to Alex Gimenez-Grau and Johan Henriksson for sharing with us some Mathematica code and some unpublished results respectively. Finally, we thank Wolfger Peelaers for comments on the manuscript.
 The work of C.M.~has received funding from the European
Union’s Horizon 2020 research and innovation programme under the Marie Skłodowska-Curie grant agreement No 754496. The work of P.F.  has received funding from the European Research Council (ERC) under the European Union’s Horizon 2020 research and innovation programme (grant agreement No 787185).

\onecolumngrid

\appendix

\section{Blockology}\label{app:blockology}

Here we discuss the conformal blocks that are relevant for the correlators considered in the body of this letter.  We shall also give explicit results for the OPE of the correlators at the first few orders in perturbation theory.  There,  one finds expressions such as $a^{(n)}_{\Delta}$,  representing OPE coefficients,  and $\gamma^{(n)}_{\Delta}$,  representing anomalous dimensions.  It is important to stress that while the symbol $\langle \cdot \rangle$ has been omitted to ease the reading,  all such quantities are always meant as averages over degenerate operators,  in the sense discussed in the body of this letter.

\subsection{The conformal blocks for $\langle \mathcal{D}_1\,\mathcal{D}_1\,\mathcal{D}_1\,\mathcal{D}_1\rangle$}\label{sec:blocks1111}

Superconformal symmetry puts strong constraints on the correlation functions of half-BPS operators $\mathcal{D}_k$.  Not only the four-point functions between all members of the supermultiplet are determined by those of the superconformal primaries, but the latter are subject to constraints.  The latter are elegantly formulated and solved in superspace,  where one has a coordinate $t$ on the line,  $\mathfrak{so}(5)\simeq \mathfrak{sp}(4)$ null polarization vectors $Y$ and fermionic coordinates.  The constraints imposed by superconformal invariance can be solved in terms of a number $\mathsf{f}$ and a function $f(\chi)$,  as follows:
\begin{align}\label{1111}
\frac{\langle \mathcal{D}_{1} \mathcal{D}_{1} \mathcal{D}_{1} \mathcal{D}_{1}\rangle}{\langle \mathcal{D}_{1} \mathcal{D}_{1}\rangle \langle \mathcal{D}_{1} \mathcal{D}_{1}\rangle}\,=
\mathsf{f}\,\sX+\mathbb{D}\ffcorrshort(\chi)\,,
\end{align}
where in terms of the spacetime and R-symmetry cross ratios
\begin{align}
\chi=\frac{t_{12}t_{34}}{t_{13} t_{24}}\,, \quad
\myzeta_1\, \myzeta_2=\frac{(Y_1\cdot Y_2)\,(Y_3\cdot Y_4)}{(Y_1\cdot Y_3)\,(Y_2\cdot Y_4)}\,,
\quad
(1-\myzeta_1)\,(1-\myzeta_2)=\frac{(Y_1\cdot Y_4)\,(Y_2\cdot Y_3)}{(Y_1\cdot Y_3)\,(Y_2\cdot Y_4)}\
\end{align}
one has $\sX \equiv \frac{\chi^{2}}{\myzeta_{1} \myzeta_{2}}$.  We have used $t_{ij}=t_i-t_j$.  We also have the differential operator $\mathbb{D}$, given by
\begin{align}
\mathbb{D}=v_1+v_2- v_1v_2\,\chi^2\partial_{\chi}\,,
\quad
v_i=\chi^{-1}-\myzeta_{i}^{-1}\,.
\end{align}

The conformal blocks for $\langle \mathcal{D}_1\,\mathcal{D}_1\,\mathcal{D}_1\,\mathcal{D}_1\rangle$ were already considered in \cite{Liendo:2018ukf}. We write them here for completeness, giving the results in terms of the constant $\mathsf{f}$ ``reduced correlator'' $\ffcorrshort(\chi)$ introduced in \eqref{1111}.  We also recall the $\mathcal{D}_1\times \mathcal{D}_1$ OPE, which reads
\begin{align}
\mathcal{D}_1\times \mathcal{D}_1=\mathcal{I}\oplus \mathcal{D}_2\oplus \mathcal{L}^{\Delta}_{0,[0,0]}\,.
\end{align}
The conformal blocks $\gshort_{\mathcal{O}}(\chi)$ for the exchange of the three operators on the right hand side are listed below.
\begin{itemize}
\item For the identity operator $\mathcal{I}$:
\begin{align}
\mathsf{f}_{\mathcal{I}}=1, \qquad
\gshort_{\mathcal{I}}(\chi)=\chi\,.
\end{align}
\item For the short supermultiplet $\mathcal{D}_2$:
\begin{align}
\mathsf{f}_{\mathcal{D}_2}=1, \qquad
\gshort_{\mathcal{D}_2}(\chi)=\chi\,\left(1-{}_2F_1(1,2,4;\chi)\right)\,.
\end{align}
\item For a generic long supermultiplet $\mathcal{L}^{\Delta}_{0,[0,0]}$:
\begin{align}\label{longblocks1111}
\mathsf{f}_{\mathcal{L}^{\Delta}_{0,[0,0]}}=0, \qquad
\gshort_{\mathcal{L}^{\Delta}_{0,[0,0]}}(\chi)=\frac{\chi^{\Delta+1}}{1-\Delta}\,{}_2F_1(\Delta+1,\Delta+2,2(\Delta+2);\chi)\,.
\end{align}
\end{itemize}

\subsection{The conformal blocks for $\langle \mathcal{D}_1\,\mathcal{D}_1\,\mathcal{D}_2\,\mathcal{L}^{\DeltaExt}_{0,[0,0]}\rangle$}\label{sec:112L}

For this correlator we use the following convention:
\begin{align}
\langle \mathcal{D}_1\,\mathcal{D}_1\,\mathcal{D}_2\,\mathcal{L}^{\DeltaExt}_{0,[0,0]}\rangle=\frac{(Y_1\cdot Y_3)(Y_2\cdot Y_3)}{t_{13}^2\,t_{23}^2}\left(\frac{t_{12}}{t_{14}\,t_{24}}\right)^{\DeltaExt}\,\ffcorrlong(\chi)\,.
\end{align}
Here we can expand in two different OPE channels. The first is the $s$-channel, corresponding to $\chi \to 0$: in this case the OPE $\mathcal{D}_1\times \mathcal{D}_1$ can be read off from the previous case, while we should also consider
\begin{align}
\mathcal{D}_2\times \mathcal{L}^{\DeltaExt}_{0,[0,0]}=\mathcal{D}_2\oplus  \mathcal{L}^{\Delta}_{0,[0,0]}+\dots\,\,\,.
\end{align}
Here the dots contain terms that do not contribute to the correlator we are considering here. Furthermore, although the three-point function with the superconformal primary of the multiplet $\mathcal{L}^{\Delta}_{0,[0,0]}$ vanishes, there is a non-zero three point function with its $Q^4$ descendant in the $R$-symmetry representation $[0,2]$.
The conformal blocks for these two operators are
\begin{itemize}
\item For the short supermultiplet $\mathcal{D}_2$:
\begin{align}
\glong_{\mathcal{D}_2}(\chi)=\left(\frac{1-\chi}{\chi}\right)^{\DeltaExt}\,{}_2F_1(\DeltaExt,2,4;\chi)\,,
\end{align}
\item For the long supermultiplet $\mathcal{L}^{\Delta}_{0,[0,0]}$:
\begin{align}
\glong_{\mathcal{L}^{\Delta}_{0,[0,0]}}(\chi)=\left(\frac{1-\chi}{\chi}\right)^{\DeltaExt}\,\chi^{\Delta}\,{}_2F_1(\Delta+\DeltaExt,\Delta+2,2(\Delta+2);\chi)\,.
\end{align}
\end{itemize}
The second OPE limit that we consider is the $u$-channel, corresponding to $\chi\to \infty$: in this case we have the two OPEs
\begin{align}
\begin{split}
\mathcal{D}_1\times \mathcal{D}_2&=\mathcal{D}_1\oplus \mathcal{D}_3\oplus \mathcal{L}^{\Delta}_{0,[0,1]}\,,\\
\mathcal{D}_1\times \mathcal{L}^{\DeltaExt}_{0,[0,0]}&=\mathcal{D}_1\oplus \mathcal{L}^{\Delta}_{0,[0,1]}\,.
\end{split}
\end{align}
The conformal blocks for the operators appearing in both OPEs are given by
\begin{itemize}
\item  For the short supermultiplet $\mathcal{D}_1$:
\begin{align}
\glong_{\mathcal{D}_1}=\left(1-\chi^{-1}\right)^{\DeltaExt}\,,
\end{align}
\item For the long supermultiplet $\mathcal{L}^{\Delta}_{0,[0,1]}$:
\begin{align}
\glong_{\mathcal{L}^{\Delta}_{0,[0,1]}}=\left(1-\chi^{-1}\right)^{\DeltaExt}\,\chi^{1-\Delta}\,{}_2F_1(\Delta-1,\Delta+\DeltaExt+3,2(\Delta+2);\chi^{-1})\,.
\end{align}
\end{itemize}

\subsection{Conformal blocks for $\langle \mathcal{D}_1\,\mathcal{D}_1\,\mathcal{D}_1\,\mathcal{D}_1\rangle$ in perturbation theory}

Consider the reduced correlator $\gshort(\chi)$ for the four-point function $\langle \mathcal{D}_1\,\mathcal{D}_1\,\mathcal{D}_1\,\mathcal{D}_1\rangle$ and its conformal blocks decomposition
\begin{align}\label{genericblocks}
\ffcorrshort(\chi)=\gshort_{\mathcal{I}}(\chi)+a_{\mathcal{D}_2}\,\gshort_{\mathcal{D}_2}(\chi)+\sum_{\mathcal{L}_{0,[0,0]}}a_{\mathcal{L}_{0,[0,0]}}\,\gshort_{\Delta_{\mathcal{L}_{0,[0,0]}}}(\chi)\,,
\end{align}
where $\gshort_{\Delta_{\mathcal{L}_{0,[0,0]}}}(\chi)$ are the relevant conformal blocks, while $a_{\mathcal{L}_{0,[0,0]}}$ are (squared) OPE coefficients. The case we are interested in is a strongly coupled theory, with perturbative parameter $1/\sqrt{\lambda}$ (and large $\lambda$). We then expand all the ingredients appearing in \eqref{genericblocks} perturbatively, namely
\begin{align}\label{perturbativedata}
\begin{split}
\ffcorrshort(\chi)&=
\ffcorrshort^{(0)}(\chi)
+\frac{1}{\lambda^{1/2}}\,\ffcorrshort^{(1)}(\chi)
+\frac{1}{\lambda}\,\ffcorrshort^{(2)}(\chi)
+\frac{1}{\lambda^{3/2}}\,\ffcorrshort^{(3)}(\chi)+\frac{1}{\lambda^{2}}\,\ffcorrshort^{(4)}(\chi)+\dots\,\,\,,\\
\Delta_{\mathcal{L}_{0,[0,0]}}&=
\Delta
+\frac{1}{\lambda^{1/2}}\,\gamma^{(1)}_{\Delta}
+\frac{1}{\lambda}\,\gamma^{(2)}_{\Delta}
+\frac{1}{\lambda^{3/2}}\,\gamma^{(3)}_{\Delta}
+\frac{1}{\lambda^{2}}\,\gamma^{(4)}_{\Delta}+\dots\,\,\,,\\
a_{\mathcal{L}_{0,[0,0]}}&=
a^{(0)}_{\Delta}
+\frac{1}{\lambda^{1/2}}\,a^{(1)}_{\Delta}
+\frac{1}{\lambda}\,a^{(2)}_{\Delta}
+\frac{1}{\lambda^{3/2}}\,a^{(3)}_{\Delta}
+\frac{1}{\lambda^{2}}\,a^{(4)}_{\Delta}+\dots\,\,\,.
\end{split}
\end{align}
Now consider the blocks for long operators, given in \eqref{longblocks1111}. As well known, when expanding the conformal dimension of such operators perturbatively, the over factor of $\chi^{\Delta}$ produces powers of $\log\chi$, the highest of which can be reconstructed at each order from the CFT data at previous orders \cite{Aharony:2016dwx}. In order to display the powers of $\log\chi$ in the small $\chi$ expansion explicitly at each order, we introduce
\begin{align}
\gshort^{(\ell)}_{\Delta}(\chi)=\chi^{\Delta} \left(\partial_{\Delta}\right)^{\ell}\chi^{-\Delta}\, \gshort_{\Delta}(\chi)\,,
\end{align} 
where $\gshort_{\Delta}(\chi)$ are conformal blocks for long operators when their dimension is taken have its free theory value.
The result is
\begin{align}\label{short0}
\ffcorrshort^{(0)}(\chi)=\gshort_{\mathcal{I}}(\chi)+a^{(0)}_{\mathcal{D}_2}\,\gshort_{\mathcal{D}_2}(\chi)+\sum_{\Delta}a^{(0)}_{\Delta}\,\gshort_{\Delta}(\chi)\,,
\end{align}
\begin{align}\label{short1}
\ffcorrshort^{(1)}(\chi)=\sum_{\Delta}\left[a^{(0)}_{\Delta}\,\gamma^{(1)}_{\Delta}\,f_{\Delta}(\chi)\right]\,\log\chi\,+\,a^{(1)}_{\mathcal{D}_2}\,\gshort_{\mathcal{D}_2}(\chi)\,+\,\sum_{\Delta}\left[a^{(1)}_{\Delta}\,\gshort_{\Delta}(\chi)+a^{(0)}_{\Delta}\,\gamma^{(1)}_{\Delta}\,\gshort^{(1)}_{\Delta}(\chi)\right]\,,
\end{align}
\begin{align}\label{short2}
\begin{split}
\ffcorrshort^{(2)}(\chi)&=\sum_{\Delta}\left[\frac{1}{2}a^{(0)}_{\Delta}\,\left(\gamma^{(1)}_{\Delta}\right)^2\,\gshort_{\Delta}(\chi)\right]\,\log^2\chi\,+\,\sum_{\Delta}\left[\left(a^{(0)}_{\Delta}\gamma^{(2)}_{\Delta}+a^{(1)}_{\Delta}\gamma^{(1)}_{\Delta}\right)\,\gshort_{\Delta}(\chi)+a^{(0)}_{\Delta}\,\left(\gamma^{(1)}_{\Delta}\right)^2\,\gshort^{(1)}_{\Delta}(\chi)\right]\,\log\chi\\
&+\,a^{(2)}_{\mathcal{D}_2}\,\gshort_{\mathcal{D}_2}(\chi)\,+\sum_{\Delta}\left[a^{(2)}_{\Delta}\,\gshort_{\Delta}(\chi)+\left(a^{(0)}_{\Delta}\gamma^{(2)}_{\Delta}+a^{(1)}_{\Delta}\gamma^{(1)}_{\Delta}\right)\,\gshort^{(1)}_{\Delta}(\chi)+\frac{1}{2}a^{(0)}_{\Delta}\,\left(\gamma^{(1)}_{\Delta}\right)^2\,\gshort^{(2)}_{\Delta}(\chi)\right]\,,
\end{split}
\end{align}
\begin{align}\label{short3}
\begin{split}
\ffcorrshort^{(3)}(\chi)&=\sum_{\Delta}\left[\frac{1}{6}a^{(0)}_{\Delta}\,\left(\gamma^{(1)}_{\Delta}\right)^3\,\gshort_{\Delta}(\chi)\right]\,\log^3\chi\\
&+
\sum_{\Delta}\left[\left(a^{(0)}_{\Delta}\,\gamma^{(1)}_{\Delta}\,\gamma^{(2)}_{\Delta}+\frac{1}{2}a^{(1)}_{\Delta}\,\left(\gamma^{(1)}_{\Delta}\right)^2\right)\,\gshort_{\Delta}(\chi)+\frac{1}{2}a^{(0)}_{\Delta}\,\left(\gamma^{(1)}_{\Delta}\right)^3\,\gshort^{(1)}_{\Delta}(\chi)\right]\,\log^2\chi\\
&+\sum_{\Delta}\left[\left(a^{(0)}_{\Delta}\,\gamma^{(3)}_{\Delta}+a^{(1)}_{\Delta}\,\gamma^{(2)}_{\Delta}+a^{(2)}_{\Delta}\,\gamma^{(1)}_{\Delta}\right)\,\gshort_{\Delta}(\chi)+\left(2\,a^{(0)}_{\Delta}\,\gamma^{(1)}_{\Delta}\,\gamma^{(2)}_{\Delta}+a^{(1)}_{\Delta}\,\left(\gamma^{(1)}_{\Delta}\right)^2\right)\,\gshort^{(1)}_{\Delta}(\chi)\right.\\
&\hspace{1.1cm}\left.+\frac{1}{2}a^{(0)}_{\Delta}\,\left(\gamma^{(1)}_{\Delta}\right)^3\,\gshort^{(2)}_{\Delta}(\chi)\right]\,\log\chi\\
&+\,a^{(3)}_{\mathcal{D}_2}\,\gshort_{\mathcal{D}_2}(\chi)\,+\sum_{\Delta}\left[ a^{(3)}_{\Delta}\,\gshort_{\Delta}(\chi)+\left(a^{(0)}_{\Delta}\,\gamma^{(3)}_{\Delta}+a^{(1)}_{\Delta}\,\gamma^{(2)}_{\Delta}+a^{(2)}_{\Delta}\,\gamma^{(1)}_{\Delta}\right)\,\gshort^{(1)}_{\Delta}(\chi)
\right.\\ &\hspace{1.1cm}\left. +
\left(a^{(0)}_{\Delta}\,\gamma^{(1)}_{\Delta}\,\gamma^{(2)}_{\Delta}+\frac{1}{2}a^{(1)}_{\Delta}\,\left(\gamma^{(1)}_{\Delta}\right)^2\right)\,\gshort^{(2)}_{\Delta}(\chi)+\frac{1}{6}a^{(0)}_{\Delta}\,\left(\gamma^{(1)}_{\Delta}\right)^3\,\gshort^{(3)}_{\Delta}(\chi)\right]\,,
\end{split}
\end{align}
\begin{align}\label{short4}
\begin{split}
\ffcorrshort^{(4)}(\chi)&=\sum_{\Delta}\left[\frac{1}{24}a^{(0)}_{\Delta}\,\left(\gamma^{(1)}_{\Delta}\right)^4\,\gshort_{\Delta}(\chi)\right]\,\log^4\chi\\
&+
\sum_{\Delta}\left[\left(\frac{1}{2}a^{(0)}_{\Delta}\,\left(\gamma^{(1)}_{\Delta}\right)^2\,\gamma^{(2)}_{\Delta}+\frac{1}{6}a^{(1)}_{\Delta}\,\left(\gamma^{(1)}_{\Delta}\right)^3\right)\,\gshort_{\Delta}(\chi)+\frac{1}{6}a^{(0)}_{\Delta}\,\left(\gamma^{(1)}_{\Delta}\right)^4\,\gshort^{(1)}_{\Delta}(\chi)\right]\,\log^3\chi\\
&+\sum_{\Delta}\left[\left(a^{(0)}_{\Delta}\,\gamma^{(1)}_{\Delta}\,\gamma^{(3)}_{\Delta}+\frac{1}{2}a^{(0)}_{\Delta}\,\left(\gamma^{(2)}_{\Delta}\right)^2+a^{(1)}_{\Delta}\,\gamma^{(1)}_{\Delta}\,\gamma^{(2)}_{\Delta}+\frac{1}{2}a^{(2)}_{\Delta}\,\left(\gamma^{(1)}_{\Delta}\right)^2\right)\,\gshort_{\Delta}(\chi)\right.\\
&\hspace{1.1cm}\left. +
\left(\frac{1}{2}a^{(1)}_{\Delta}\,\left(\gamma^{(1)}_{\Delta}\right)^3+\frac{3}{2}a^{(0)}_{\Delta}\,\left(\gamma^{(1)}_{\Delta}\right)^2\,\gamma^{(2)}_{\Delta}\right)\,\gshort^{(1)}_{\Delta}(\chi)+\frac{1}{4}a^{(0)}_{\Delta}\,\left(\gamma^{(1)}_{\Delta}\right)^4\,\gshort^{(2)}_{\Delta}(\chi)\right]\,\log^2\chi\\
&+\sum_{\Delta}\left[\left(a^{(0)}_{\Delta}\,\gamma^{(4)}_{\Delta}+a^{(1)}_{\Delta}\,\gamma^{(3)}_{\Delta}+a^{(2)}_{\Delta}\,\gamma^{(2)}_{\Delta}+a^{(3)}_{\Delta}\,\gamma^{(1)}_{\Delta}\right)\,\gshort_{\Delta}(\chi)\right.\\&\hspace{1.1cm}\left. +
 \left(a^{(0)}_{\Delta}\,\left(\gamma^{(2)}_{\Delta}\right)^2+2\,a^{(0)}_{\Delta}\,\gamma^{(1)}_{\Delta}\,\gamma^{(3)}_{\Delta}+2\,a^{(1)}_{\Delta}\,\gamma^{(1)}_{\Delta}\,\gamma^{(2)}_{\Delta}+(a^{(2)}_{\Delta}\,\left(\gamma^{(1)}_{\Delta}\right)^2\right)\,\gshort^{(1)}_{\Delta}(\chi)\right.\\
 &\hspace{1.1cm}\left. +\left(\frac{3}{2}a^{(0)}_{\Delta}\,\left(\gamma^{(1)}_{\Delta}\right)^2\,\gamma^{(2)}_{\Delta}+\frac{1}{2}a^{(1)}_{\Delta}\,\left(\gamma^{(1)}_{\Delta}\right)^3\right)\,\gshort^{(2)}_{\Delta}(\chi)+\frac{1}{24}a^{(0)}_{\Delta}\,\left(\gamma^{(1)}_{\Delta}\right)^4\,\gshort^{(3)}_{\Delta}(\chi)\right]\,\log\chi\\
 &+\,a^{(4)}_{\mathcal{D}_2}\,\gshort_{\mathcal{D}_2}(\chi)\,++\sum_{\Delta}\left[ a^{(4)}_{\Delta}\,\gshort_{\Delta}(\chi)+ \left(a^{(0)}_{\Delta}\,\gamma^{(4)}_{\Delta}+a^{(1)}_{\Delta}\,\gamma^{(3)}_{\Delta}+a^{(2)}_{\Delta}\,\gamma^{(2)}_{\Delta}+a^{(3)}_{\Delta}\,\gamma^{(1)}_{\Delta}\right)\,\gshort^{(1)}_{\Delta}(\chi) \right.\\
 &\hspace{1.1cm}\left. +\left(\frac{1}{2}a^{(0)}_{\Delta}\,\left(\gamma^{(1)}_{\Delta}\right)^2\,\gamma^{(2)}_{\Delta}+\frac{1}{6}a^{(1)}_{\Delta}\,\left(\gamma^{(1)}_{\Delta}\right)^3\right)\,\gshort^{(3)}_{\Delta}(\chi)+\frac{1}{24}a^{(0)}_{\Delta}\,\left(\gamma^{(1)}_{\Delta}\right)^4\,\gshort^{(4)}_{\Delta}(\chi)
 \right].
\end{split}
\end{align}
As discussed in \cite{Liendo:2018ukf}, the squared OPE coefficient $a_{\mathcal{D}_2}$ for the exchange of the short operator $\mathcal{D}_2$ is known from supersymmetric localization \cite{Pestun:2007rz,Giombi:2009ds,Giombi:2018qox}.  Its perturbative expansion to the first orders is
\begin{align}
\mu^2_{\mathcal{D}_2}=2-\frac{3}{\lambda^{1/2}}+\frac{45}{8\,\lambda^{3/2}}+\frac{45}{4\,\lambda^2}+\dots\,\,\,.
\end{align}

\subsection{Conformal blocks for $\langle \mathcal{D}_1\,\mathcal{D}_1\,\mathcal{D}_2\,\mathcal{L}^{\DeltaExt}_{0,[0,0]}\rangle$ in perturbation theory}

For this correlator we consider perturbative expansions analogous to those of eq. \eqref{perturbativedata}, and here we write the OPE for the first three perturbative orders in the two channels considered in section \ref{sec:112L}. Here we are bootstrapping a correlator with a long operator, which acquires an anomalous dimension in perturbation theory:
\begin{align}
\DeltaExt=\DeltaExt^{(0)}+\frac{1}{\lambda^{1/2}}\gamma^{(1)}_{\ext}+\frac{1}{\lambda}\gamma^{(2)}_{\ext}+\dots\,\,\,.
\end{align}
When expanding the OPE in the $s$-channel,  we introduce
\begin{align}
\glong_{\mathcal{O}}^{(\ell_1,\ell_2)}(\chi)=\chi^{\Delta-\DeltaExt}\left(\partial_{\Delta}\right)^{\ell_1}\left(\partial_{\DeltaExt}\right)^{\ell_2}\,\chi^{-\Delta+\DeltaExt}\glong_{\mathcal{O}}(\chi)\,,
\end{align}
and we have the expansions
\begin{align}
\begin{split}
\ffcorrlong^{(0)}(\chi)&=a^{(0)}_{\mathcal{D}_2}\,\glong_{\mathcal{D}_2}(\chi)+\sum_{\Delta}a^{(0)}_{\Delta}\,\glong_{\Delta}(\chi)\,,
\end{split}
\end{align}
\begin{align}
\begin{split}
\ffcorrlong^{(1)}(\chi)&=\left[-a^{(0)}_{\mathcal{D}_2}\,\gamma^{(1)}_{\ext}\,\glong_{\mathcal{D}_2}(\chi)+\sum_{\Delta}a^{(0)}_{\Delta}\,\left(\gamma^{(1)}_{\Delta}-\gamma^{(1)}_{\ext}\right)\,\glong_{\Delta}(\chi)\right]\,\log\chi\\
&+\left[
a^{(1)}_{\mathcal{D}_2}\,\glong_{\mathcal{D}_2}(\chi)+a^{(0)}_{\mathcal{D}_2}\,\gamma^{(1)}_{\ext}\,\glong^{(0,1)}_{\mathcal{D}_2}(\chi)
+\sum_{\Delta}\left(
a^{(1)}_{\Delta}\,\glong_{\Delta}(\chi)
+a^{(0)}_{\Delta}\,\gamma^{(1)}_{\Delta}\,\glong^{(1,0)}_{\Delta}(\chi)
+a^{(0)}_{\Delta}\,\gamma^{(1)}_{\ext}\,\glong^{(0,1)}_{\Delta}(\chi)
\right)\right]\,,
\end{split}
\end{align}
\begin{align}\label{112L_secondorder}
\begin{split}
\ffcorrlong^{(2)}(\chi)&=\left[\frac{1}{2}a^{(0)}_{\mathcal{D}_2}\,\left(\gamma^{(1)}_{\ext}\right)^2\,\glong_{\mathcal{D}_2}(\chi)+\sum_{\Delta}\frac{1}{2}a^{(0)}_{\Delta}\,\left(\gamma^{(1)}_{\Delta}-\gamma^{(1)}_{\ext}\right)^2\,\glong_{\Delta}(\chi)\right]\,\log^2\chi\\
&+\left[-\left(a^{(1)}_{\mathcal{D}_2}\,\gamma^{(1)}_{\ext}+a^{(0)}_{\mathcal{D}_2}\,\gamma^{(2)}_{\ext}\right)\,\glong_{\mathcal{D}_2}(\chi)-a^{(0)}_{\mathcal{D}_2}\,\left(\gamma^{(1)}_{\ext}\right)^2\,\glong^{(0,1)}_{\mathcal{D}_2}(\chi)\right.\\
 &\hspace{0.5cm}\left. +\sum_{\Delta}\left(a^{(0)}_{\Delta}\left(\gamma^{(2)}_{\Delta}-\gamma^{(2)}_{\ext}\right)+a^{(1)}_{\Delta}\left(\gamma^{(1)}_{\Delta}-\gamma^{(1)}_{\ext}\right)\right)\,\glong_{\Delta}(\chi)\right.\\
 &\hspace{0.5cm}\left. + \sum_{\Delta}a^{(0)}_{\Delta}\,\gamma^{(1)}_{\Delta}\,\left(\gamma^{(1)}_{\Delta}-\gamma^{(1)}_{\ext}\right)\,\glong^{(1,0)}_{\Delta}(\chi)+ \sum_{\Delta}a^{(0)}_{\Delta}\,\gamma^{(1)}_{\ext}\,\left(\gamma^{(1)}_{\Delta}-\gamma^{(1)}_{\ext}\right)\,\glong^{(0,1)}_{\Delta}(\chi)
\right]\,\log\chi\\
&+\left[a^{(2)}_{\mathcal{D}_2}\,\glong_{\mathcal{D}_2}(\chi)+\left(a^{(0)}_{\mathcal{D}_2}\,\gamma^{(2)}_{\ext}+a^{(1)}_{\mathcal{D}_2}\,\gamma^{(1)}_{\ext}\right)\,\glong^{(0,1)}_{\mathcal{D}_2}(\chi)+\frac{1}{2}a^{(0)}_{\mathcal{D}_2}\,\left(\gamma^{(1)}_{\ext}\right)^2\,\glong^{(0,2)}_{\mathcal{D}_2}(\chi)\right.\\
 &\hspace{0.5cm}\left. +\sum_{\Delta}a^{(2)}_{\Delta}\,\glong_{\Delta}(\chi)+\sum_{\Delta}\left(a^{(0)}_{\Delta}\,\gamma^{(2)}_{\Delta}+a^{(1)}_{\Delta}\,\gamma^{(1)}_{\Delta}\right)\,\glong^{(1,0)}_{\Delta}(\chi)+\sum_{\Delta}\left(a^{(0)}_{\Delta}\,\gamma^{(2)}_{\ext}+a^{(1)}_{\Delta}\,\gamma^{(1)}_{\ext}\right)\,\glong^{(0,1)}_{\Delta}(\chi)\right.\\
 &\hspace{0.5cm}\left. +\sum_{\Delta}\frac{1}{2}a^{(0)}_{\Delta}\left(\gamma^{(1)}_{\Delta}\right)^2\,\glong^{(2,0)}_{\Delta}(\chi)
 +\sum_{\Delta}\frac{1}{2}a^{(0)}_{\Delta}\left(\gamma^{(1)}_{\ext}\right)^2\,\glong^{(0,2)}_{\Delta}(\chi)
+\sum_{\Delta}a^{(0)}_{\Delta}\,\gamma^{(1)}_{\Delta}\,\gamma^{(1)}_{\ext}\,\glong^{(1,1)}_{\Delta}(\chi)\right]\,.
\end{split}
\end{align}
An important comment at this point concerns the part of $\ffcorrlong^{(2)}(\chi)$ that is proportional to $\log\chi$ in the OPE above,  which in the main text we referred to as $\ffcorrlong^{(2)}_{\log}(\chi)$.  Note that all the combinations of CFT data appearing there are known from the results at previous orders,  except for terms containing $a^{(0)}_{\Delta}\,\gamma^{(2)}_{\Delta}$ and $a^{(0)}_{\Delta}\,\gamma^{(2)}_{\mathtt{ext}}$.  The former should be better interpreted as 
\begin{align}
\langle a^{(0)}_{\Delta}\,\gamma^{(2)}_{\Delta}\rangle_{112\mathcal{O}_{\mathtt{ext}}}\,,
\end{align}
which is the quantity we are interested in computing when we look at this correlators.  As described in the main text, the strategy is that of computing enough of these averages from the known values of $\Gamma^{(2)}_{\Delta,2 \to 4}$, so that $\ffcorrlong^{(2)}_{\log}(\chi)$ is completely fixed when the other constraints are also taken into account. 

The story is quite different for $a^{(0)}_{\Delta}\,\gamma^{(2)}_{\mathtt{ext}}$. In particular, one might think that it is necessary to consider, as an external operator $\mathcal{L}_{0,[0,0]}^{\DeltaExt}$ an eigenstate of dilatation operator. In that case, the quantity $\gamma^{(2)}_{\ext}$ that appears in the OPE of $\ffcorrlong^{(2)}(\chi)$ would be one of the eigenvalues of the dilatation operator at second order, $\Gamma^{(2)}_{\DeltaExt}$. What we want to stress here is that this, although correct in principle, is not necessary. One is actually free to chose any linear combination of eigenstates as an external operator, thus avoiding the diagonalization process, and in that case the $\gamma^{(2)}_{\ext}$ appearing on the second and third line of \eqref{112L_secondorder} should be interpreted as follows:
\begin{align}\label{gamma2ext_112L}
\begin{split}
\ffcorrlong^{(2)}_i(\chi)\supset\gamma^{(2)}_{\ext}\,\left[a^{(0)}_{\mathcal{D}_2}\,\glong_{\mathcal{D}_2}(\chi)+\sum_{\Delta}a^{(0)}_{\Delta}\,\glong_{\Delta}(\chi)\right] \longrightarrow \sum_{j}\left(\Gamma^{(2)}_{\DeltaExt}\right)_{ij}\,\ffcorrlong^{(0)}_j(\chi)\,.
\end{split}
\end{align}
Here we have chosen an arbitrary basis for the operators with $\omega=\{\DeltaExt,0,[0,0]\}$ in the free theory, labelled by indices $i$ and $j$. $\ffcorrlong^{(n)}_i(\chi)$ then refers to the $n$-th order correlator with the $i$-th operator in the chosen basis. 

The way we bootstrap the $\log\chi$ part of $\ffcorrlong^{(2)}(\chi)$ is to input the averages $\langle a^{(0)}_{\Delta}\,\gamma^{(2)}_{\Delta}\rangle$ for $2\le \Delta \le \DeltaExt+2$. This fixes that all the other averages completely, without need to specify which basis of operators we are using: the explicit matrix entries $\left(\Gamma^{(2)}_{\DeltaExt}\right)_{ij}$ of eq. \eqref{gamma2ext_112L} drop out completely from all equations once one inputs enough data.

In the $u$-channel, we introduce
\begin{align}
\glong_{\mathcal{O}}^{(\ell_1,\ell_2)}(\chi)=\chi^{-\Delta}\left(\partial_{\Delta}\right)^{\ell_1}\left(\partial_{\DeltaExt}\right)^{\ell_2}\,\chi^{\Delta}\glong_{\mathcal{O}}(\chi)\,,
\end{align}
and we have the expansions
\begin{align}
\ffcorrlong^{(0)}(\chi)=a^{(0)}_{\mathcal{D}_1}\,\glong_{\mathcal{D}_1}(\chi)+\sum_{\Delta}a^{(0)}_{\Delta}\,\glong_{\Delta}(\chi)\,,
\end{align}
\begin{align}
\begin{split}
\ffcorrlong^{(1)}(\chi)&=\left[-\gamma^{(1)}_{\ext}\,\sum_{\Delta}a^{(0)}_{\Delta}\,\glong_{\Delta}(\chi)\right]\,\log\chi+\left[a^{(1)}_{\mathcal{D}_1}\,\glong_{\mathcal{D}_1}(\chi)+a^{(0)}_{\mathcal{D}_1}\,\gamma^{(1)}_{\ext}\,\glong^{(0,1)}_{\mathcal{D}_1}(\chi)\right]\\
 & +\sum_{\Delta}\left[a^{(1)}_{\Delta}\,\glong_{\Delta}(\chi)+a^{(0)}_{\Delta}\,\gamma^{(1)}_{\Delta}\,\glong^{(1,0)}_{\Delta}(\chi)+a^{(0)}_{\Delta}\,\gamma^{(1)}_{\ext}\,\glong^{(0,1)}_{\Delta}(\chi)\right]\,,
\end{split}
\end{align}
\begin{align}
\begin{split}
\ffcorrlong^{(2)}(\chi)&=\sum_{\Delta}\left[\frac{1}{2}a^{(0)}_{\delta}\,\left(\gamma^{(1)}_{\Delta}\right)^2\,\glong_{\Delta}(\chi)\right]\,\log^2\chi\\
&+\sum_{\Delta}\left[-\left(a^{(0)}_{\Delta}\,\gamma^{(2)}_{\Delta}+a^{(1)}_{\Delta}\,\gamma^{(1)}_{\Delta}\right)\,\glong_{\Delta}(\chi)-a^{(0)}_{\Delta}\,\left(\gamma^{(1)}_{\Delta}\right)^2\,\glong^{(1,0)}_{\Delta}(\chi)-a^{(0)}_{\Delta}\,\gamma^{(1)}_{\Delta}\,\gamma^{(1)}_{\ext}\,\glong^{(0,1)}_{\Delta}(\chi)\right]\,\log\chi\\
&+\left[a^{(2)}_{\mathcal{D}_1}\,\glong_{\mathcal{D}_1}(\chi)+\left(a^{(0)}_{\mathcal{D}_1}\,\gamma^{(2)}_{\ext}+a^{(1)}_{\mathcal{D}_1}\,\gamma^{(1)}_{\ext}\right)\,\glong^{(0,1)}_{\mathcal{D}_1}(\chi)+\frac{1}{2}a^{(0)}_{\mathcal{D}_1}\,\left(\gamma^{(1)}_{\ext}\right)^2\,\glong^{(0,2)}_{\mathcal{D}_1}(\chi)\right]\\
&+\sum_{\Delta}\left[
a^{(0)}_{\Delta}\,\glong_{\Delta}(\chi)+\left(a^{(0)}_{\Delta}\,\gamma^{(2)}_{\Delta}+a^{(1)}_{\Delta}\,\gamma^{(1)}_{\Delta}\right)\,\glong^{(1,0)}_{\Delta}(\chi)+\left(a^{(0)}_{\Delta}\,\gamma^{(2)}_{\ext}+a^{(1)}_{\Delta}\,\gamma^{(1)}_{\ext}\right)\,\glong^{(0,1)}_{\Delta}(\chi)\right.\\
&\hspace{1.1cm}\left.
+\frac{1}{2}a^{(0)}_{\Delta}\,\left(\gamma^{(1)}_{\Delta}\right)^2\,\glong^{(2,0)}_{\Delta}(\chi)
+a^{(0)}_{\Delta}\,\gamma^{(1)}_{\ext}\,\gamma^{(1)}_{\Delta}\,\glong^{(1,1)}_{\Delta}(\chi)
+\frac{1}{2}a^{(0)}_{\Delta}\,\left(\gamma^{(1)}_{\ext}\right)^2\,\glong^{(0,2)}_{\Delta}(\chi)
\right]\,.
\end{split}
\end{align}
Note that here $\gamma_{\ext}^{(2)}$ only enters the part of $\glong^{(2)}(\chi)$ which does not contain any power of $\log\chi$: in our bootstrap method we do not reconstruct that part as it is not necessary to compute entries of the anomalous dimension matrix at second order, $\Gamma^{(2)}_{\Delta}$.

\section{The basis of transcendental functions}\label{app:basis}

The analytic structure of four-point functions in a 1d CFT in terms of the unique cross-ratio $\chi$ was discussed in detail in \cite{Mazac:2018qmi}: in the complex $\chi$-plane, such correlators are analytic functions with branch points at $\chi=0,1$. Thus, when making an ansatz in terms of harmonic polylogarithms,  only this type of singularities is allowed: such functions can be obtained using the symbol map as ``words'' made of the two ``letters''  $\chi$ and $1-\chi$.  The number of such words is $2^{\mathtt{t}}$,  which then gives us the dimension of our basis for weight $\mathtt{t}$.  Using functional relations between polylogarithms,  we choose the following basis for $\mathtt{t}\le4$:

\begin{itemize}
\item $\mathtt{t}=0$: $\{1\}$.
\item $\mathtt{t}=1$: $\{\log(\chi),\,\log(1-\chi)\}$.
\item $\mathtt{t}=2$: $\{\log^2(\chi),\,\log(\chi)\log(1-\chi),\,\log^2(1-\chi),\,\text{Li}_2(\chi)\}$.
\item $\mathtt{t}=3$: $\{\log^3(\chi),\,\log^2(\chi)\log(1-\chi),\,\log(\chi)\log^2(1-\chi),\,\log^3(\chi),\,\text{Li}_2(\chi)\log(\chi),\,\text{Li}_2(\chi)\log(1-\chi),$\\
${}\hspace{1.2cm} \text{Li}_3(\chi),\,\text{Li}_3\left(\frac{\chi}{\chi-1}\right)\}$.
\item $\mathtt{t}=4$: $\{\log^4(\chi),\,\log^3(\chi)\log(1-\chi),\,\log(\chi)^2\log^2(1-\chi),\,\log(\chi)\log^3(1-\chi),\,\log^4(\chi),\,\text{Li}_2(\chi)\log^2(\chi),$\\
${}\hspace{1.2cm} \,\text{Li}_2(\chi)\log(\chi)\log(1-\chi),\,\,\text{Li}_2(\chi)\log^2(1-\chi),\,\text{Li}_3(\chi)\log(\chi),\,\text{Li}_3(\chi)\log(1-\chi),$\\
${}\hspace{1.25cm}\text{Li}_3\left(\frac{\chi}{\chi-1}\right)\log(\chi),\,\text{Li}_3\left(\frac{\chi}{\chi-1}\right)\log(1-\chi),\,\text{Li}_4(\chi),\,\text{Li}_4(1-\chi),\,\text{Li}_4\left(\frac{\chi}{\chi-1}\right) \}$.
\end{itemize}

A comment is also in order regarding the rational functions that multiply the polylogarithms in our ansatz.  Given the analytic structure of correlators in 1d CFTs discussed above,  the only allowed singularities for these rational functions are poles at $\chi=0,1$.  It follows from this that they must be given by polynomials in $\chi$,  divided by products of powers of $\chi$ and $(1-\chi)$.

In the body of the letter we have claimed that a correct basis of harmonic polylogarithms for the problem at hand has transcendentality $\mathtt{t}=\ell$ at order $\ell$.  One could take this as a working assumption,  justified by the fact that it allows to find solutions to crossing compatible with the expectations at each perturbative order that we have considered.  However,  it can be also argued by looking at the highest powers of $\log\chi$ that appear in the OPE around $\chi=0$.  For simplicity we focus on $\langle \mathcal{D}_1\,\mathcal{D}_1\,\mathcal{D}_1\,\mathcal{D}_1\rangle$,  where from equations (\ref{short0}-\ref{short4}) it can be argued that
\begin{align}
\ffcorrshort^{(\ell)}_{\log^{\ell}}=\frac{1}{\ell!}\sum_{\Delta}a^{(0)}_{\Delta}\,\left(\gamma^{(1)}_{\Delta}\right)^{\ell}\,\gshort_{\Delta}(\chi)\,.
\end{align}
One can easily to this sum for every fixed $\ell$,  and the result is always a polynomial in $\chi$,  so that the contribution to the transcendentality due to this piece of the correlator is entirely due to $\log^{\ell}(\chi)$.  Similar considerations apply to all $\ffcorrshort^{(\ell)}_{\log^{k}}$ with $2\le k\le \ell$: all the sums that can be performed using CFT data from previous orders give a function of maximal transcendentality $\ell$ at order $\ell$.

\section{Anomalous dimensions in $\langle \mathcal{D}_1\,\mathcal{D}_1\,\mathcal{D}_1\,\mathcal{D}_1\rangle$}\label{app:1111}

For the reader's convenience we write here the anomalous dimensions for the long operators $\mathcal{L}_{0,[0,0]}^{\Delta}$ averaged in the $\langle \mathcal{D}_1\,\mathcal{D}_1\,\mathcal{D}_1\,\mathcal{D}_1\rangle$ correlator, up to fourth order.  The results for the correlators and the OPE coefficients can be found in a notebook included with the submission.
We have:
\begin{align}
\begin{split}
\langle\gamma^{(1)}_{\Delta}\rangle &=\gamma^{(1)}_{\Delta}=-\frac{1}{2}\,\JJ\,,\\
\langle\gamma^{(2)}_{\Delta}\rangle &=\frac{1}{8} \,\JJ\, \left(\frac{(\Delta -1) \left(4 \Delta ^2+11 \Delta +8\right)}{\JJ+2}+4 H_{\Delta }\right)\,,\\
\langle\gamma^{(3)}_{\Delta}\rangle &=\frac{\JJ}{4}\,H^{(2)}_{\Delta}-\frac{\JJ}{2}\left(H_{\Delta}\right)^2+\frac{1}{4} \,\JJ\, \left(\JJ-2\right)\,S_{-2}(\Delta)-\frac{(\Delta -1) \left(2 \Delta ^4+12 \Delta ^3+24 \Delta ^2+21 \Delta +12\right)}{2 \,(\JJ+2)}\,H_{\Delta}\\
&-\frac{\Delta  \left(5 \Delta ^5+23 \Delta ^4+17 \Delta ^3-45 \Delta ^2-71 \Delta -21\right)}{8\,(\JJ+2)}\,.
\end{split}
\end{align}
As in the main text we have introduced the quantity
\begin{align}
\JJ=\Delta(\Delta+3)\,,
\end{align}
corresponding to the Casimir eigenvalue for operators with in the representation $\omega=\{\Delta,0,[0,0]\}$ of $\mathfrak{osp}(4^*|4)$. We have also introduced the harmonic sum
\begin{align}\label{harmonicsum}
S_{-2}(\Delta)=\sum_{n=1}^{\Delta}\frac{(-1)^n}{n^2}=\frac{(-1)^{\Delta}}{4}\left(H^{(2)}_{\Delta/2}-H^{(2)}_{(\Delta-1)/2}\right)-\frac{1}{2}\zeta(2)\,,
\end{align}
where we used the generalized harmonic numbers
\begin{align}
H^{(m)}_{n}=\sum_{k=1}^{n}\frac{1}{k^m}\,,
\end{align}
with $H_n\equiv H^{(1)}_n$.  It might appear that \eqref{harmonicsum} is not analytic in $\Delta$ because of the factor $(-1)^{\Delta}$. However, recall that in the expressions above $\Delta$ is always an even number, so that such factor is effectively 1 in the case of interest.  We note that the one-loop averaged anomalous dimension $\langle\gamma^{(2)}_{\Delta}\rangle$ was computed in \cite{Liendo:2018ukf} with a similar bootstrap computation to the one presented here.  In \cite{Liendo:2018ukf},  the problem of mixing was ignored,  so that the results obtained there at one loop could have been,  a priori,  incorrect.  However,  as noticed in the main text the operators mixing is not lifted at tree level,  which makes one has $\langle\left(\gamma^{(1)}_{\Delta}\right)^2\rangle=\left(\gamma^{(1)}_{\Delta}\right)^2$,  so that the results found here agree with those of \cite{Liendo:2018ukf}.

Let us also give here more explicitly on of the main results of our work,  namely the expression of 
 \begin{align}\label{gammasquared}
\langle a^{(0)}_{\Delta}\left(\gamma^{(2)}_{\Delta}\right)^2\rangle=\langle a^{(0)}_{\Delta}\rangle\,\left[\left(\Gamma^{(2)}_{\Delta,2\to 2}\right)^2+\delta \Gamma_{\mathtt{sq}}^{(2)}(\Delta)\right]\,,
\end{align}
where the average is taken on the $\langle \mathcal{D}_1\,\mathcal{D}_1\,\mathcal{D}_1\,\mathcal{D}_1\rangle$ correlator.  While from the free theory one finds
\begin{align}
\langle a^{(0)}_{\Delta}\rangle=\frac{\sqrt{\pi }\, 2^{-2 \Delta -1}\, (\Delta -1)\, \Gamma [\Delta +3]}{\Gamma \left[\Delta +\tfrac{3}{2}\right]}\,,
\end{align}
the first term on the right hand side of \eqref{gammasquared} was computed already in \cite{Liendo:2018ukf},  and is simply
\begin{align}
\Gamma^{(2)}_{\Delta,2\to 2}=\langle\gamma^{(2)}_{\Delta}\rangle\,,
\end{align}
namely the average given above.  The non-trivial result is the second term,  which is found to be
\begin{align}
\begin{split}
\delta\Gamma_{\mathtt{sq}}^{(2)}(\Delta)\,=\,\JJ \left(\frac{(\JJ-2)}{2}\,S_{-2}(\Delta)+\frac{3\,\JJ-4}{8}\,H_{\Delta}^2\right)-\frac{p_1(\Delta)}{(\Delta+1)(\Delta+2)}\,H_{\Delta}+\frac{p_2(\Delta)}{(\Delta+2)}\,,
\end{split}
\end{align}
with the polynomials $p_{1,2}(\Delta)$ given by
\begin{align}
\begin{split}
p_1(\Delta)&=\frac{1}{4}\left(4 \Delta ^6+33 \Delta ^5+92 \Delta ^4+97 \Delta ^3+23 \Delta ^2-21 \Delta -36\right)\,,\\
p_2(\Delta)&=\frac{\Delta^2}{32}\left(29 \Delta ^3+200 \Delta ^2+431 \Delta +300\right)\,.
\end{split}
\end{align}

Let us now discuss a couple of explicit examples of the anomalous dimension mixing matrix $\Gamma^{(2)}_{\Delta}$,  namely the cases of $\Delta=4,6$.  For $\Delta=4$, there only two degenerate operators,  one of length two and one of length four, which are schematically of the form ``$\partial^2 \phi^2$'' and ``$\phi^4$'' respectively.  The matrix $\Gamma^{(2)}_{\Delta=4}$ is then $2\times 2$,  and we find the results
\begin{equation}
\Gamma^{(2)}_{\Delta=4}=
\begin{pmatrix}
\Gamma^{(2)}_{\Delta=4,2\to 2} & \Gamma^{(2)}_{\Delta=4,2\to 4}\\
\Gamma^{(2)}_{\Delta=4,2\to 4} & \Gamma^{(2)}_{\Delta=4,4\to 4}
\end{pmatrix}=
\begin{pmatrix}
\frac{2093}{30} & 7 \sqrt{\frac{5}{2}}\\
7\sqrt{\frac{5}{2}} & \frac{203}{4}
\end{pmatrix}\,\,.
\end{equation}
For $\Delta=6$,  the degeneracy space has dimension four,  including one operator of length two (``$\partial^4 \phi^2$''),  two operators of length four (both of the schematic form ``$\partial^2 \phi^4$'') and one operator of length six (``$\phi^6$'').  Making an arbitrary choice of basis in the space of length four operators,  which we shall not discuss here explicitly,  one can write the $4\times 4$ matrix $\Gamma^{(2)}_{\Delta=6}$ as
\begin{equation}
\Gamma^{(2)}_{\Delta=6}=
\begin{pmatrix}
   \Gamma^{(2)}_{\Delta=4,2\to 2} & \Gamma^{(2)}_{\Delta=4,2\to 4} &  \Gamma^{(2)}_{\Delta=4,2\to 6} \\ 
\Gamma^{(2)}_{\Delta=4,2\to 4} & \Gamma^{(2)}_{\Delta=4,4\to 4}& \Gamma^{(2)}_{\Delta=4,4\to 6}\\ 
\Gamma^{(2)}_{\Delta=4,6\to 2} & \Gamma^{(2)}_{\Delta=4,6\to 4} & \Gamma^{(2)}_{\Delta=4,6\to 6}
    \end{pmatrix}
=
\begin{pmatrix}
 \frac{110619}{560} & \frac{9 \sqrt{165}}{8} & \frac{3 \sqrt{429}}{2} &0\\ 
\frac{9 \sqrt{165}}{8} &  \frac{3541}{24} &   \frac{28}{3} \sqrt{\frac{13}{5}} & \sqrt{\frac{1365}{2}}\\
\frac{3 \sqrt{429}}{2} & \frac{28}{3} \sqrt{\frac{13}{5}} & \frac{20323}{120} & \sqrt{42} \\
0 & \sqrt{\frac{1365}{2}} & \sqrt{42} & \frac{1035}{8}
  \end{pmatrix}\,.
\end{equation}
Note that $\Gamma^{(2)}_{\Delta=4,2\to 6}=0$ is not a coincidence, but rather a consequence of the structure of the dilatation operator at order $\mathcal{O}(1/\lambda)$,  which can connect only operators whose length differs at most by two units.

\section{First order anomalous dimension matrix}\label{app:treegammas}

In the following we will sketch a derivation of equation (17).  Let us denote by $\mathbb{D}^{(1)}$ the first order correction to the dilatation operator and assume that it has the following properties:
\begin{itemize}
\item[1.] $\mathbb{D}^{(1)}$ commutes with the action of $\mathfrak{osp}(4^*|4)$ defined in the free theory. \item[2.] $\mathbb{D}^{(1)}$ acts within each $\mathcal{H}_L$ defined in equation (9), in other words it does not mix states of different lengths.
 We denote the corresponding blocks by 
$\mathbb{D}^{(1)}_L$.
\item[3.] The first order correction to the dilatation operator takes the form   
\begin{equation}
\mathbb{D}^{(1)}_L=\sum_{1\leq i < j \leq L}\,\mathbf{d}_{ij}\,,
\qquad 
\mathbf{d}:=\mathbb{D}^{(1)}_{L=2}\,,
\end{equation}
where we use the standard notation in which the indices $ij$ in $\mathbf{d}_{ij}$ indicate on which factors in the product $\mathcal{H}_L$ the operator $\mathbf{d}$ is acting on. Notice that $\mathbb{D}^{(1)}_L$ is permutation invariant so its action on $\mathcal{H}_L$  is well defined.
\item[4.] The operator $\mathbb{D}^{(1)}_{L=2}$ is proportional to the quadratic Casimir operator acting on $\mathcal{H}_2$
\begin{equation}
\mathbb{D}^{(1)}_{L=2}=-\tfrac{1}{2}\,\widehat{\mathfrak{C}}_2(J^{(12)})=: -J^{(1)}\cdot J^{(2)} \,,
\qquad
\widehat{\mathfrak{C}}_2(\mathfrak{J})= k^{x y} \mathfrak{J}_x\,\mathfrak{J}_y\,,
\end{equation}
where $ \mathfrak{J}_\alpha$  is a basis $\mathfrak{osp}(4^*|4)$, $k^{xy}$ is the inverse Killing form, $J^{(12)}:=J^{(1)}+J^{(2)}$ and  each $J^{(i)}$ corresponds to the representation of the single letter $\mathbb{V}_{\Phi}\simeq \mathcal{D}_1$. We also introduced the notation $J^{(1)}\cdot J^{(2)}$.
\end{itemize}
Equation (17) follows from the properties 1.--\,4. after noticing that the single letter representation has zero Casimir, namely
 $\widehat{\mathfrak{C}}_2(J^{(i)})=0$, more explicitly
\begin{equation}
\label{D1Loperator}
\mathbb{D}^{(1)}_L=
-\sum_{1\leq i < j \leq L}\,J^{(i)}\cdot J^{(j)}\,=\,
-\tfrac{1}{2}\, \sum_{1\leq i, j \leq L}J^{(i)}\cdot J^{(j)}
\,=\,
-\tfrac{1}{2}\,\widehat{\mathfrak{C}}_2(J^{(12\dots L)})\,,
\end{equation}
where $J^{(12\dots L)}=J^{(1)}+J^{(2)}+\dots +J^{(L)}$. To compare \eqref{D1Loperator} with (17) in the paper we recall that  $\mathfrak{C}_2(\mathcal{R}_{\mathcal{O}})$ are by definition  the eigenvalues of $\widehat{\mathfrak{C}}_2(J^{(12\dots L)})$.
This concludes the proof of (17) under the assumptions 1.--\,4. \\

   Let us comment on the origin of these properties in our case. Property 1 follows from similar arguments to the ones used in \cite{Beisert:2003zd}.
   Properties 2--4 all follow from the fact that at first order in perturbation theory any correlator is derived from the quartic interaction which is encoded in $\langle \mathcal{D}_1\mathcal{D}_1\mathcal{D}_1\mathcal{D}_1\rangle$. The latter can be either observed directly from the structure of the field theory in AdS of from the perturbative bootstrap, see \cite{Longversion} for more details.
   More precisely, to derive properties 2 and 3 one needs to further observe that a quartic vertex (like that in eq. (2.9) of \cite{Giombi:2017cqn}) could in principle produce, together with an operator $\mathbf{d}: \mathbb{V}_{\Phi} \otimes \mathbb{V}_{\Phi} \rightarrow \mathbb{V}_{\Phi} \otimes \mathbb{V}_{\Phi}$ also terms of the form $\mathbb{V}_{\Phi}  \rightarrow \mathbb{V}_{\Phi}^{\otimes 3}$ and  $\mathbb{V}_{\Phi}^{\otimes 3} \rightarrow \mathbb{V}_{\Phi}$.
   These are not compatible with the first property. Finally, property $4$ follows from property 1 together with the fact that we know the spectrum of $\mathbb{D}^{(1)}_{L=2}$,  since it can be extracted directly from $\langle \mathcal{D}_1\mathcal{D}_1\mathcal{D}_1\mathcal{D}_1\rangle$.

\end{document}